# Should policy makers trust composite indices?
## A commentary on the pitfalls of inappropriate indices for policy formation

Matthias Kaiser[1,a], Andrew Tzer-Yeu Chen[2,b], Peter Gluckman[2,c]

[1] The Centre for the Study of the Sciences and the Humanities (SVT),
University of Bergen, Norway

[2] Koi Tū: The Centre for Informed Futures, The University of Auckland, New Zealand

[a] Corresponding Author, matthias.kaiser@uib.no

[b] andrew.chen@auckland.ac.nz   [c] pd.gluckman@auckland.ac.nz

**Abstract:**

**Background:** This paper critically discusses the use and merits of global indices, in particular, the Global Health Security Index or GHSI (Cameron et 2019) in times of an imminent crisis, like the current pandemic. The index ranked 195 countries according to their expected preparedness in case of a pandemic or other biological threat. The Covid-19 pandemic provides the background to compare each country's predicted performance from the GHSI with the actual performance. In general, there is an inverted relation between predicted versus actual performance, i.e. the predicted top performers are among those that are the worst hit. Obviously, this reflects poorly on the potential policy uses of the index in imminent crisis management.

**Methods:** The paper analyses the Global Health Security Index and identifies why it may have struggled to predict actual pandemic preparedness as evidenced by the Covid-19 pandemic. The paper also uses two different data sets, one from the Worldmeter on the spread of the Covid-19 pandemics, and the other one from the INGSA policy tracker, to make comparisons between the actual introduction of pandemic response policies and the corresponding death rate in 29 selected countries.

**Results:** This paper analyses the reasons for the poor match between prediction and reality in the index, and mentions six general observations applying to global indices in this respect. These observations are based on methodological and conceptual analyses. The level of abstraction in these global indices builds uncertainties upon uncertainties and hides implicit value assumptions which potentially removes them from the policy needs on the ground.

**Conclusions:** From the analysis, the question is raised if the policy community might have better tools for decision making in a pandemic. On the basis of data from the International Network for Government Science Advice (INGSA) Policy-making Tracker, and with backing in studies from social psychology and philosophy of science, some simple heuristics are suggested, which may be more useful than a global index.

\*\*\*

**Background**:

Why would anyone want to construct a global composite index of anything? The standard answer is that it would offer a useful tool for policy design and decision-making. In theory, the composite score is easier to understand than a complex concept such as wellbeing or sustainability because it provides a quantified measure. The next question is then: how can users know if the index is good and useful? Tentatively, we might suggest that the index has value if using it leads to better policies and decisions than would have been the case without it. However, most global composite indicators are never really put to the test, because performance is normally not directly measurable. We surmise that the case of the Global Health Security Index (GHSI) [1] may be an exception. This index was published in October 2019 after two and a half years of research, and contained a ranking of 195 countries with their associated scores indicating their preparedness for global epidemics and pandemics. The GHSI aimed to be a key resource in the "face of increasing risks of high-consequence and globally catastrophic biological events in light of major gaps in international financing for preparedness" [1]. The developers "believe that, over time, the GHS Index will spur measurable changes in national health security" and sought to "illuminate preparedness and capacity gaps to increase both political will and financing to fill them at the national and international levels." [1] It utilized 140 questions organized in 6 categories, 34 indicators and 85 sub-indicators, all constructed from open-source information. Out of a total possible score of 100, the average for these countries was 40.2, ranging between 83.5 to 16.2. Fewer than 7% of the countries are ranked as being able to effectively prevent the emergence or release of pathogens.

**Methods:**

Our guiding question was: How accurate was the GHSI? To answer this , we compared two data-sets: the Worldometer data on the sdpread of the Covid-19 pandemic and the INGSA policy tracker data, both reduced to 29 selected countries based on gross household incomes. The bulk of the paper then analyses the GHSI and identifies challenges for why it may not have been well suited for predicting performance in pandemic response, before demonstrating the power of a simple heuristic in the conclusion.

Less than half a year afer the publication of the GHSI, the novel SARS-CoV-2 coronavirus led to the Covid-19 pandemic. This gives us now the possibility to compare the previous assessment from the index with the actual performance of countries' health systems. Firstly, we can see how the GHSI ranked countries into three levels of preparedness. The United States and the United Kingdom top the index at rank 1 and 2 (scoring 83.5 and 77.9), with Sweden (72.2), South Korea (70.2) and France (68.1) with high rankings at 7, 9, and 11 respectively. Then there are countries like Germany (66.0), Spain (65.9), Norway (64.9), Italy (56.2), New Zealand (54.0) and others which are placed in the middle category of preparedness, apparently not so well prepared. Brazil (59.7, rank 22) is ranked slightly better than Singapore (58.7, rank 24). Mongolia is at least above the average with a score of 49.5, while Jamaica (29.0, rank 147) and Fiji (25.7, rank 168) are among those ranked as very poorly prepared.

**Results:**

For those who have been watching the global spread of Covid-19, the incongruities between the GHSI rankings and the case numbers in each country will be obvious. June 2020 data from the Worldometer website [https://www.worldometers.info/coronavirus/] shows the opposite of what we might expect from the GHSI rankings: The United States, United Kingdom, Sweden, and Brazil are the worst hit countries (a trend that continued later), while other countries have been surpassing expectations derived from the index: Germany, Norway, Singapore, New Zealand, and Vietnam. This is certainly also the case for Jamaica and Fiji, which have virtually eliminated Covid-19, but were ranked among the least prepared. Two of the best performing economies (in June 2020), Taiwan and Hong Kong, are not even included in the global index. We do note that there are some limitations with the use of data aggregators like Worldometer and potential issues around the reliability of testing in some countries, but we feel that these data are suitable enough for demonstrating the magnitude of the problem. Table 1 shows a comparison of 29 selected countries, chosen on the basis of gross household income. This subset was chosen as those with the highest household incomes to focus on countries where resource limitations in the health care system would not be the primary determinant. A small rank difference indicates a close estimate between GHSI and reality, while a large rank difference indicates a big gap between expectation and actual performance. Out of 178 countries, only 20 had a Worldometer ranking within 20 rank positions of their ranking on the GHSI.

[INSERT TABLE 1 HERE, SEE PAGE 11]

Quantifying the expected and actual performance data for these 29 countries invites us to look more deeply into why the differences were so large. It is not always obvious what should count as performance data – one might pick the number of infections per million inhabitants, or as we have, the number of deaths per million people. We think it is generally desirable to avoid a high death rate. What we have not done is to break down the total GHSI score into its 6 subcategories, which would somewhat complicate the picture[1]. These subcategories in the GHSI report are:

1. Prevention of the emergence or release of pathogens;
2. Early detection & reporting of epidemics of potential international concern;
3. Rapid response to and mitigation of the spread of an epidemic;
4. Sufficient & robust health system to treat the sick & protect health workers;
5. Commitments to improving national capacity, financing and adherence to norms;
6. Overall risk environment and country vulnerability to biological threats.

Any single country will score differently on these subcategories, such that the GHSI claimed to provide more detailed information on where to act in order to improve general preparedness. We surmise that this might indeed hold true for at least some of the long-term policy improvements in those countries. Yet, the total score is what is used for international comparison and ranking, and what conveys the most weight in political discussion. In late February, US President Donald Trump cited the GHSI to argue that the United States was well prepared for Covid-19, saying "the United States, we're rated No. 1" [3]. And here the discrepancy arises most clearly with actual performance.

In the subsequent section, we discuss the challenges with the GHSI, and in the conclusion present a simpler heuristic that could provide a more direct explanation for the success (or not) of pandemic responses.

**Discussion and Literature Review:**

We want to ask why was the index so wrong?

There have already been several assessments of how the GHSI fared in the light of Covid-19 ([4], [5], [6], [7]). All noted significant shortcomings, and based on data from Worldometer several noted the reversal of relationships. Razavi et al [6] questions the wisdom of the ranking system: "ranking countries based on weighted scores across indicators that are scored variably and are not directly comparable with one another is problematic". They based this assertion on their criticism that the scoring system is not consistent (some indicators score either 0 or 100, while others use the whole range), the use of weightings is arbitrary, and the inclusion of some indicators like urbanization, are of questionable validity. Chang and McAleer [7] extended the analysis of GHSI by adding other approaches to quantify mean values (adding to the arithmetic mean used in GHSI, the geometric and harmonic mean values) and saw positive potential in the GHSI, but stressed the significant differentiation in the 6 indicators: "Rapid Response and Detection and Reporting have the largest impacts" [7]. They also commented on the implicit political bias: "As part of China, Hong Kong was not included in the GHS Index as a country, while Taiwan was not included undoubtedly for political reasons" [7]. Abbey et al [5] observe the incongruity between prediction and outcome, and stress that political leadership and previous experiences with epidemics should be counted as a crucial factor for preparedness, and therefore be included in the future improvements of the GHSI.

---

[1] However, a quick glance at the rankings of the six categories can already convey that no radical changes can be expected that would bring the combined ranking closer to the Worldometer data.

With the partial exception of the Aitken et al critique, most of the criticism focusses on the technical aspects of constructing the composite index, in particular when it comes to combining sub-categories to create a total score assigned to the individual countries. Abbey et al [5] seem to find the main shortcoming in incorrect weighting and incomplete definition of the relevant categories. It is noteworthy, though, that they seem to find pragmatic policy value in various sub-categories, thereby putting aside the problem of how these were constructed from a large number of indicators and sub-indicators.

Given the extreme discrepancy between expected and actual performance for most countries, one must ask if this failure might be due to a deeper inherent weakness in the underlying concept of the index, or other contextual factors. For instance, could underperformance simply be due to political decision makers not utilizing their countries' capabilities or feeling overconfident? Could performance exceeding expectations be due to political decision makers compensating for the lack of preventative capacity through more stringent interventions, perhaps supported by geographical luck? All of this might have been a factor in the actual performance during the imminent crisis. We would, however, argue that it is wrong to solely put the blame or praise on the side of politics, when in all of these countries the decision-making was presented as evidence-based, and the GHSI purports to capture the whole range of what then seemed relevant, publicly-available facts.

It is important to note that we have little evidence that the index actually formed a key part of policy making in governments around the world, but that it is clear the index was constructed with this intention in mind. Given that the GHSI was obviously meticulously prepared, based on a wealth of data by a large group of international experts, we might even generalize the question to now ask whether the production of any such global composite indicator makes any sense at all as a strategic decision-making tool when facing a global imminent crisis. In other words, is the intention of supplying a decision-support tool via a global index in time of an imminent health crisis realistic?

**Constructing a composite index**

Initially, a composite index is simply a function of a series of underlying indicators. There are a number of steps and decisions involved in constructing good composite indicators. According to Mazziotta & Pareto [8] the following decision steps are crucial:

(i) *Defining the problem!* For some issues there may exist a clear and agreed upon definition about how the problem is to be defined, and what its constitutive elements and basic categories are. The challenge here is to have a reliable system theoretic understanding which establishes nodes and contact points with other systems. With complex problems, e.g. the grand societal challenges, this step is often a topic of scientific dispute and even conflict. In regard to threats to health security when facing pandemics, it is an ongoing discussion what factors and categories are crucial.

(ii) *Selecting a group of individual indicators!* "Ideally, indicators should be selected according to their relevance, analytical soundness, timeliness, accessibility, etc. The selection step is the result of a trade-off between possible redundancies caused by overlapping information and the risk of losing information." (ibid., 70). Again, a good insight into systemic interdependencies is necessary to make that choice. It involves judgement about the relative importance of individual indicators and avoiding those which are strongly correlated to others, since this increases redundancy. The indicators need to be measurable, i.e. they need concrete data to support them. Here, other issues arise, e.g. in regard the choice of scale of measurement (e.g. national, regional, household etc.), or the numerical units of measurement (0 and 1, or a range of values between 0 and n?). If data are insufficient in one country, one may in certain cases apply Monte Carlo methods to impute missing data.

(iii) *Normalizing the selected indicators!* Here one has to make sure that comparability is achieved between different indicators among various categories, despite different measurement dimensions. Therefore, one transforms them to simple numbers, and those can be achieved by different means like e.g. ranking, distance to a reference point etc. This often has the effect of abstracting detail away, adding uncertainty to the accuracy of the number.

(iv) *Aggregating the normalized indicators!* This step combines the chosen indicators to just one or several aggregate indices. Since they are normalized, a simple method can be mathematical addition of unit rankings. However, in complex situations one might face the need to weigh them differently, i.e. to assign weights to each indicator before they are aggregated into the final index to emphasise some indicators over others. Again, the assignment of weights is obviously a matter of subjective judgement, and we would surmise often strongly biased by chosen value-perspectives: "Different weighting techniques may be chosen, none of which is exempt by a discretionary choice" [9]. Furthermore, how are we to combine the different indicators? Most commonly this could be done by using weighted arithmetic mean values, but since this can imply that bad performance on one indicator can be countered by sufficiently good performance on others, it may not be adequate for what we want to measure in the combined index. One may then choose a weighted geometric mean, for these values. There are formal methods to reduce the dependency on mere judgement, e.g. by assigning weight according to the variability of the indicator, since the greater the variability, we assume the greater its influence on what we try to measure with the index. However, even then our delineations will depend on an overall judgement.

In the scientific literature there are a number of contributions on how to improve the performance of composite indices, for instance in the area of food security, which admittedly is a very complex issue and approached from different angles (cf. [9], [10], [11]). There is significant progress in some of the technical issues involved in the construction of these indicators.

Yet, what one easily can infer from the above description is that building a composite index is not straightforward and involves a series of subjective judgments. It also involves dealing with system uncertainties, and dealing with value-presuppositions (cf. [12, [13]). Mazziotta and Pareto [8] express this the following way:

"No universal method exists for composite indices construction. In each case their construction is much determined by the particular application, including both formal and heurist elements, and incorporate some expert knowledge on the phenomenon." ([8], 72).

One may ask then the following question: if the process to construct global composite indices is so complex and difficult, why do we have such a plethora of them around in the first place? The cynical answer would be because this is what scientists can do. The more neutral and probably more realistic answer would be because they communicate easily to decision-makers. The typical policy question might be: 'How are we doing in regard to X?' where X is a rather complex societal problem, and then the follow up would be: 'And how are we doing in regard to X compared to other countries?' If we are able to provide a simple answer, preferably expressed as a number relation like a rank, then the decision-maker regards this as highly useful, specifically when followed up by proposed measures how to improve that number. The usual problem is that X as such is not dealt with in any singular scientific silo due to its complexity, but needs to be translated and broken down into sub-problems which are conducive to the apparatus of the various scientific disciplines. This is a task involving trans-disciplinarity. Therefore, we concur that the intention of the construction of composite indices as planning tool is a valid one, but we also maintain that valuable information loss occurs on the way from the scientist to the decision-maker. The information loss is often related to the implicit uncertainties or the implicit values as they become more invisible the closer and higher one moves to the decision-maker [14]. It is this information loss and simplification of outcome which severely restricts the use of a global index as a management tool in an imminent crisis situation.

We make the following observations in this regard:

(1) Indices like the GHSI comprise of several layers of specificity, looking for measurable (quantifiable) features that are considered essential for higher-level properties. By implication, it emerges that higher-level properties are not directly measurable, and that is why one seeks to circumvent the problem by using subordinated indicators which indirectly contribute to the higher-level property. Typically, these higher-level properties are not directly measurable due to their complexity, implying the possibility of diverse emergent and unpredictable phenomena. The issue is known in the literature as the subjectivity and contentious nature of identifying the very problem in the first place, and then breaking down the problem in measurable sub-units. Some global indices have successfully identified a simple list of parameters that people agree to be essential for the issue. A case in point is the Human Development Index [15], where there has been wide agreement on the basic three individual indicators. In the case of the GHSI, they broke down biological preparedness into six categories with 34 indicators and 85 sub-indicators. However, while already the selection of the categories may be questioned, one may certainly also question the chosen indicators and sub-indicators. What comprises health security is debated and a matter of subjective judgement. Implicit in the choice are problematic questions like whether or not some indicators are substitutable or non-substitutable. Can, for instance, wide social compliance with protective measures and good risk communication reduce the number of necessary intensive care beds?

(2) The upward process from the concrete to the more abstract implies building uncertainties upon uncertainties, without the means to precisely quantify them. In global studies, large uncertainties are typically already present through differences in how base local data are registered and counted. Communicating a single index score or rank for each country masks the inherent uncertainty and volatility in the measurements. While statistical methods exist to e.g. combine uncertainty with sensitivity analysis [10], they do not eliminate deep uncertainty but simply transfer them into quantifiable units [11].

(3) Groups of properties that fall under a common concept are presumed to be uniformly linear and additive when contributing towards a common concept. This excludes local variation in response to the threat / property indexed in the study. One such variation might be cultural differences in risk communication. It also ignores interdependencies and mutual strengthening or weakening of constitutive properties. Typically one tries to circumvent this by normalization of the various indicators to make them comparable to each other. While the process of normalization can be done in various ways, it will by necessity involve some subjective judgement on systemic dependencies.

(4) Constructing a global composite index as a strategic tool in decision-making presupposes the existence of a normative benchmark for ideal states. Any such benchmark will indirectly introduce a socio-political and cultural bias that does not do justice to the diversity of viewpoints among both experts and non-experts. Countries with high compliance to imposed rules and regulations may have other needs in terms of preparedness than countries with low compliance. This point could be less important, if the users have the time to critically assess the method the index is built on, in particular the match of the underlying problem (in our case the preparedness to health threats) and the chosen indicators with their local conditions. However, in times of an imminent crisis one must assume that each crisis provides its own complex challenges, and the initial efforts are understanding the particular nature of the threat. For instance, the kind and speed of transmission, the need for protective gear (PPE), contact tracing methods, availability of intensive care units, or the possibility and accessibility of vaccination emerged gradually as assessment tasks while the current crisis of Covid-19 already was a fact.

(5) While solid and comprehensive reporting of the way a global composite indicator was constructed may shield one from some academic criticism, the fact remains that users of the index, the policy makers, will almost certainly focus on the overall performance figures as reflected in the indicator. In the current example, the authors of the GHSI stated that "national health security is fundamentally weak around the world" and that "no country is fully prepared for epidemics or pandemics, and every country has important gaps to address." [3] But that did not stop an MP in the South African National Assembly from claiming "the Global Health Security Index for 2019 Report revealed that South Africa is ranked 34 out of 195 countries... this gives confidence that the South African government through the national Department of Health is doing all within its power to strengthen its health systems to safeguard the public from the outbreak of any other form of infections" [16]. The key utility of a global index is ultimately being able to rely on the overall performance ranking.

(6) Breaking down a politically important property into its sub-indicators and other elements always runs the risk of leaving out system interactions and interdependencies which are not routinely assessed in specific academic disciplines. For instance, while the coronavirus pandemic in most countries activated epidemiologists and virologists, in some countries it also activated psychologists, economists, social scientists and philosophers. Thus, there is bias already in the framing of the issue, and alternative framings that would emerge in a transdisciplinary approach are seldom considered.

It is by no means surprising that scientific systematization will always include uncertainties and will never be completely objective. Facts and values are intertwined in science for policy. The framework of post-normal science [12, 13] has stressed this for a long time. It has also been pointed out in areas outside of health, that composite indices may be misleading and may hide important information in some of its elements. Giampietro and Saltelli [17] have raised this issue in regard the global ecological footprint (EF), and this has spurred a number of reactions [18, 19, 20]. We also note other composite indices that rely on poor proxies [8,11], such as the OECD Better Life Index [21] or the university rankings [22]. The danger is that policy decisions based on flawed indices and rankings are likely to be equally flawed.

**What might be better?**

Let us put the question the other way round: could one make sensible and scientifically informed policies without these global indicators or index? With the experience of Covid-19 fresh in our minds, we would venture that good pandemic policies (leaving out the other issues for the time being) could be based on and started with sensible data presentation and some simple heuristics rather than over-stated modelling with its inherent limitations. One key to effective control of the pandemic was acting preventatively at an early stage, and implementing counter-measures like widespread testing, lockdown and closing of the borders [23]. Taiwan is one of the best examples in this respect: noting the rapid rise of infections in neighboring China in late December 2019, it implemented wide-spread testing among incoming people, and set in motion a National Health Command Center. It soon closed its borders, quarantined all cases, and propagated face masks rapidly. They certainly did not find any advice in the GHSI since they were not included in the first place. Early detection and reaction were the key to controlling the pandemic in many countries, and they showed success. Laissez-faire attitudes like in the UK, Sweden, or the USA proved fatal. A UN report has this key message: "Act decisively and early to prevent the further spread or quickly suppress the transmission of Covid-19 and save lives" [24]. Other writers have already noted that simplicity may be a better guide than getting lost in the complexities: "An imperative to prioritize simplicity over complexity is at the core of social health" [25].

Furthermore, in all modelling it is a widely recognized trade-off between precision and complexity. Complex models are seen as more accurate, while simple ones are seen as more general with a lack of detail causing systematic bias in predictions—but adding detail to a model does not guarantee an increase in reliability unless the added processes are essential, well understood, and reliably estimated [26]. O'Neill's conjecture was that there may be an optimal balance between model complexity and model error ([26], p. 70). In our case this implies that adding to the complexity of the basic categories in the GHSI may actually increase model error rather than decrease it. This is also the background for the recommendations in [27]. To quote this article:

> "Complexity can be the enemy of relevance. Most modelers are aware that there is a tradeoff between the usefulness of a model and the breadth it tries to capture. But many are seduced by the idea of adding complexity in an attempt to capture reality more accurately. As modelers incorporate more phenomena, a model might fit better to the training data, but at a cost. Its predictions typically become less accurate. As more parameters are added, the uncertainty builds up (the uncertainty cascade effect), and the error could increase to the point at which predictions become useless." ([27] p. 483).

Now, we want to stress that we are concerned with meeting an immediate health crisis and we ask whether or not an index like the GHSI can be regarded as a useful addition in our toolbox to manage that crisis. Above, we have already claimed that as a matter of fact it was not useful, and certainly not a precise predictor. But the real underlying question we want to ask is if we are looking in the wrong toolbox altogether. As with all tools, the utility of the tool depends on its intended use. Therefore, we do not argue that there is no use for a composite index like the GHSI, since we might assume that it could have good uses as a tool in the design of long-term strategies in our health policies. What we, however, claim, is that policy in an imminent crisis like a pandemic is ill advised if looking at the composite index as a guide to crisis management. This does not imply that science cannot contribute to crisis management, rather the opposite: it is highly useful if the right information comes in the right format and the right doses at the right times. It only implies that scientific advice may take other roads to policy than a global composite index. One issue might be to resist the temptation to providing numbers, i.e., quantification, when the problem is still poorly understood.

"Quantification can backfire. Excessive regard for producing numbers can push a discipline away from being roughly right towards being precisely wrong." ([27] p. 484).

One needs to recognize the immediate needs of decision-makers facing an immediate crisis. Obviously, a decision-maker tries to come up with robust decisions, and robust decisions are typically about a set of different future scenarios under deep uncertainty, and guided by varying criteria for robustness, as e.g. a decision-maker may change from an optimist to a pessimist strategy or the other way round [28, 29]. Decision-makers need to engage in a learning process as the crisis unfolds, and thus apply an adaptive management approach [30] through the different phases of the crisis. The availability of a risk register may be a decision-support tool during this process. As the science-policy interface is known to be full of pitfalls, institutionalized brokerage may be an important support [31, 32 33, 34, 35], aiming at synthesis when information is sparce and beset with deep uncertainty. Heuristics may be more important than formal tools aiming at characterizing the whole complexity of the issue at hand. As Todd and Gigerenzer [36] observe, simple heuristics often perform comparably or better than more advanced algorithms, and they add a much-desired simplicity which leads to more robust decisions. This point does not invalidate some other uses of quantified formal indices or models, but it stresses that the scientific input needs to meet the constraints and context of the decision situation in a crisis.

In such a setting, the availability of reliable data on the emergence of the risk is typically providing a good input for decision heuristics. We illustrate this by reference to our INGSA policy tracker.

**Conclusions:**

Based on our INGSA Policy-Making Tracker project [https://www.ingsa.org/covid/policymaking-tracker-landing/], we have been able to analyze the interventions taken by over 120 countries and when they took place. From this data, we have seen two particular patterns so far. In East Asian countries such as Japan and South Korea, governments took swift action to increase the supply of PPE and face masks, and began public education campaigns at a very early stage – at least 14 days before the third death, avoiding the need for harsh restrictions. In some other developed countries, lockdown measures such as limiting gathering sizes, closing schools, limiting non-essential movement, and closing borders, were implemented well before the threat of Covid-19 spiraled out of control. The countries that have fared the worst in terms of deaths per million people waited longer before implementing similar policies, as shown in Table 2. A similar analytical approach was taken by journalists at POLITICO when comparing lockdown measures across Europe [37]. It should be noted that a number of countries with fragmented state-level responses were not included in the Table below.  A pattern like this obviously does not capture the complexity of the situation – for example, it does not incorporate the implementation or enforcement of these policies, nor does it distinguish between policies at different scales. And there are always exceptions, but the simple heuristic of acting preventatively and quickly is visible in the data.

[INSERT TABLE 2 HERE, SEE PAGE 12]

From the above, we hypothesize that there are some decision steps that may serve as useful and simple heuristics for performing well in a pandemic:

1) Recognize the threat to your country and the need for a response early;
2) Agree on a broad and transparent societal basis what response strategy is most acceptable for your country, such as elimination of the virus within your borders, "flattening the curve", or keeping the occurrence of infections below a predefined damage threshold;

3) Fill the chosen response strategy with a combination of practical measures appropriate to the epidemiological, economical, and socio-cultural circumstances, and monitor their effectiveness as the crisis unfolds;
4) Adjust, or change your chosen strategy according to the predefined goals and current data, preferably by inclusion of various sources of expertise and good communication with society.

Through further analysis of the INGSA Policy-Making Tracker data, we aim to understand the role of timing of interventions, the levels of responses, and understand the sources of evidence and justification behind these approaches. This will allow us to categorize the different types of strategies taken by governments around the world, and identify different styles of leadership and ideological underpinnings. It is noteworthy that the most obvious control mechanism successfully used, border closure, was not seen by the WHO as a key response, yet border closure was key to elimination in island states where early closure was perhaps easier. However, there is likely no global silver bullet to avoid or contain an emerging pandemic. Realizing the diversity of values along with epidemiological, economic and cultural considerations into a robust strategy is also a bridge to societal compliance.

**Keep it simple**

We conclude with the hypothesis that in order to be prepared for pandemics or manage other imminent crises, we may not need more sophistication in the construction of global composite indicators. In these situations, these indices may not be as useful as they purport to be. We may actually do without them to a large extent, and could learn from the past instead, recognizing the power of simple heuristics that make sense for the context and point us in the right direction. We may still consult global indices with a critical spirit and deep understanding of their inherent assumptions when working out long-term strategies for improvements in various sectors of policy. Yet, in a pandemic like the current one we see simple heuristics and adaptive management as key to robust policies.

**List of Abbreviations**:

GHSI: Global Health Security Index

INGSA: International Network for Government Science Advice

WHO: World Health Organisation

**Declarations:**

**Ethics approval and consent to participate**: Due to the nature of the study and the data sets, no ethics approval was necessary.

**Adherence to national and international regulations**: Not applicable.

**Consent for publication**: Not applicable.

**Availability of data and materials**: The datasets generated during and/or analysed during the current study are available in the INGSA COVID-19 Evidence-to-Policy Tracker repository, https://www.ingsa.org/covid/policymaking-tracker-landing/; the Worldometer COVID-19 Coronavirus Pandemic repository, https://www.worldometers.info/coronavirus/?; and the 2019 Global Health Security Index, https://www.ghsindex.org/#l-section--map.

**Competing interests:** The authors declare that they have no competing interests.

**Funding:** MK gratefully acknowledges funding from the Norman Barry Foundation, enabling his stay in NZ. The research is also supported by the International Network for Government Science Advice (INGSA).

**Author's contributions**: All authors, MK, AC, and PG have substantially contributed to the final version of the paper, and have read and accepted it. MK has been lead-author, supplying the first drafts of several versions of it, PD has edited and discussed these versions, and AC has supplied the data analysis and text edits. All authors fulfill the criteria for authorship as set out in the editorial policy of the journal.

**Acknowledgements**: Matthias Kaiser gratefully acknowledges funding from the Norman Barry Foundation which enabled his stay at Koi Tū – The Centre for Informed Futures at the University of Auckland, and enabled his contribution to this paper. All authors are grateful for the contribution of Jackson Blewden, and for helpful discussion with Tatjana Buklijas, Kristiann Allen, and Anne Bardsley.

TABLE 1: A selection of 29 countries with the highest gross household income [2]. The rankings are reassigned out of 178, based on the countries in both GHSI and Worldometer datasets as of 16 June 2020. The table is sorted by the rank difference (Worldometer Rank based on the number of deaths per million people - GHSI).

| Country | GHSI Rank | GHS Score | Worldometer Rank | Deaths/mil | **Rank Difference** |
|---|---|---|---|---|---|
| United Kingdom | 2 | 77.9 | 176 | 618 | **174** |
| United States | 1 | 83.5 | 171 | 359 | **170** |
| Netherlands | 3 | 75.6 | 170 | 354 | **167** |
| Sweden | 7 | 72.1 | 173 | 489 | **166** |
| Canada | 5 | 75.3 | 166 | 218 | **161** |
| France | 11 | 68.2 | 172 | 453 | **161** |
| Spain | 15 | 65.9 | 175 | 580 | **160** |
| Belgium | 19 | 61.0 | 178 | 834 | **159** |
| Denmark | 8 | 70.4 | 155 | 103 | **147** |
| Ireland | 23 | 59.0 | 169 | 346 | **146** |
| Germany | 14 | 66.0 | 158 | 106 | **144** |
| Finland | 10 | 68.7 | 149 | 59 | **139** |
| Slovenia | 12 | 67.2 | 143 | 52 | **131** |
| Austria | 26 | 58.5 | 152 | 76 | **126** |
| Norway | 16 | 64.6 | 139 | 45 | **123** |
| Luxembourg | 67 | 43.8 | 162 | 176 | **95** |
| Kuwait | 59 | 46.1 | 150 | 71 | **91** |
| Czech Republic | 42 | 52.0 | 131 | 31 | **89** |
| Saudi Arabia | 48 | 49.3 | 129 | 30 | **81** |
| Israel | 54 | 47.3 | 133 | 33 | **79** |
| South Korea | 9 | 70.2 | 66 | 5 | **57** |
| Japan | 21 | 59.8 | 78 | 7 | **57** |
| Australia | 4 | 75.5 | 59 | 4 | **55** |
| Qatar | 82 | 41.2 | 123 | 28 | **41** |
| Bahrain | 88 | 39.4 | 124 | 28 | **36** |
| Singapore | 24 | 58.7 | 57 | 4 | **33** |
| New Zealand | 36 | 54.0 | 62 | 4 | **26** |
| Malta | 97 | 37.3 | 114 | 20 | **17** |

TABLE 2: A selection of countries and policies, highlighting the number of days each policy was implemented since the third death from Covid-19. The table is sorted by the Worldometer Rank based on the number of deaths per million people. N/A indicates that this policy had not yet been implemented in that country as of 16 June 2020.

| | | | | Days since third death when policy intervention was implemented | | | | |
|---|---|---|---|---|---|---|---|---|
| Country | Worldometer Rank (deaths/mil) | Deaths per million (16 June 2020) | Date of Third Death from Covid-19 | Gathering Sizes Limited or Events Cancelled | Schools Closed | Non-essential Shops Closed | Non-essential Movement Banned | Borders Closed |
| Belgium | 178 | 834 | 03/11 | 2 | 2 | 7 | 7 | 9 |
| United Kingdom | 176 | 618 | 03/08 | 9 | 15 | 13 | 15 | N/A |
| Spain | 175 | 580 | 03/06 | 4 | 5 | 8 | 8 | 10 |
| Sweden | 173 | 489 | 03/16 | -5 | 1 | N/A | N/A | 3 |
| Netherlands | 170 | 354 | 03/08 | 4 | 8 | N/A | 15 | 10 |
| Denmark | 155 | 103 | 03/15 | -9 | -2 | 3 | -2 | -1 |
| Kuwait | 150 | 71 | 04/14 | -36 | -44 | -31 | -23 | -32 |
| Slovenia | 143 | 52 | 03/23 | -16 | -7 | -7 | -3 | N/A |
| Norway | 139 | 45 | 03/18 | -6 | -6 | N/A | N/A | -2 |
| Israel | 133 | 33 | 03/24 | -13 | -12 | -9 | -5 | -6 |
| Czech Republic | 131 | 31 | 03/25 | -15 | -15 | -11 | -9 | -9 |
| Saudi Arabia | 129 | 30 | 03/26 | -14 | -17 | -10 | -3 | -11 |
| Bahrain | 124 | 28 | 03/24 | -7 | -28 | 2 | N/A | -6 |
| Malta | 114 | 20 | 04/11 | -30 | -30 | -19 | N/A | -21 |
| New Zealand | 62 | 4 | 04/11 | -26 | -19 | -19 | -17 | -23 |
| Singapore | 57 | 4 | 03/29 | -3 | 10 | 9 | N/A | -6 |